%% file: main.tex
\documentclass[a4paper,ninepoint]{article}

\usepackage{itgspeech2023}    
\usepackage{times}            
\usepackage[english]{babel}   
\usepackage[ansinew]{inputenc}
\usepackage[T1]{fontenc}      
\usepackage[sort&compress,numbers]{natbib}	
\usepackage{amsmath,amssymb}
\usepackage{graphicx}
\usepackage[colorlinks=false,pdfborder={0 0 0}]{hyperref}
\usepackage{siunitx}
\usepackage[acronym]{glossaries}

\usepackage{multicol}

\usepackage{xspace}
\usepackage{cleveref}
\usepackage{multirow}
\usepackage{pgfplots}
\pgfplotsset{compat=1.18}
\usepackage[normalem]{ulem}
\usepackage{booktabs}
\usepackage{microtype}
\usepackage{hyperref}
\usepackage[normalem]{ulem}

\usepackage{adjustbox}
\usepackage[skip=5pt]{caption}

\usepackage{tikz}

\newcommand{\LibriCSS}{\textit{LibriCSS}\xspace}
\newcommand{\NTLibri}{\textit{LibriWASN}\xspace}
\newcommand{\NTLibriTwo}{\textit{LibriWASN$^{200}$}\xspace}
\newcommand{\NTLibriEight}{\textit{LibriWASN$^{800}$}\xspace}
\newcommand{\asnupbTwo}{\textit{asnupb2}\xspace}
\newcommand{\asnupbFour}{\textit{asnupb4}\xspace}
\newcommand{\asnupbSeven}{\textit{asnupb7}\xspace}

\newcommand{\SysA}{Sys-1\xspace}
\newcommand{\SysB}{Sys-2\xspace}
\newcommand{\SysC}{Sys-3\xspace}
\newcommand{\SysD}{Sys-4\xspace}

\input{glossaries}

\title{LibriWASN:  A Data Set for Meeting Separation, Diarization, and Recognition with Asynchronous Recording Devices}

\author{Joerg Schmalenstroeer, Tobias Gburrek, Reinhold Haeb-Umbach}

\address{Paderborn University, Department of Communications Engineering\\
  Email: \texttt{\{schmalen, gburrek, haeb\}@nt.uni-paderborn.de}}

\begin{document}
\setlength{\textfloatsep}{7pt}
\setlength{\intextsep}{7pt}
\setlength{\intextsep}{3pt}
\setlength{\abovedisplayskip}{2pt}
\setlength{\abovedisplayshortskip}{3pt}
\setlength{\belowdisplayskip}{3pt}
\setlength{\belowdisplayshortskip}{3pt}
\clubpenalty = 10000
\widowpenalty = 10000
\displaywidowpenalty = 10000

\maketitle

\begin{abstract}
We present LibriWASN, a data set whose design follows closely the LibriCSS meeting recognition data set, with the marked difference that the data is recorded with devices that are randomly positioned on a meeting table and whose sampling clocks are not synchronized.
Nine different devices, five smartphones with a single recording channel and four microphone arrays, are used to record a total of 29 channels. Other than that, the data set follows closely the LibriCSS design: the same LibriSpeech sentences are played back from eight loudspeakers arranged around a meeting table and the data is organized in subsets with different percentages of speech overlap. LibriWASN is meant as a test set for clock synchronization algorithms, meeting separation, diarization and transcription systems on ad-hoc wireless acoustic sensor networks. Due to its similarity to LibriCSS, meeting transcription systems developed for the former can readily be tested on LibriWASN. The data set is recorded in two different rooms and is complemented with ground-truth diarization information of who speaks when.
\end{abstract}

\section{Introduction}


\Gls{CSS} refers to the task of mapping a continuous incoming data stream consisting of the speech of an arbitrary number of, possibly concurrently active, speakers to a fixed number of output channels in such a way that there is no longer speech overlap on any of the output channels. It is designed to be a source separating preprocessing step of a meeting recognition system. \LibriCSS is a well-known publicly available data set for evaluating and comparing meeting transcription systems~\cite{20_Chen_LibriCSS}. It is structured in sets with different percentages of speech overlap, i.e., the amount of time when two speakers are concurrently active. In each set eight loudspeakers at fixed positions play back clean recordings from the \textit{LibriSpeech} corpus~\cite{Libri15} while a synchronous seven-channel circular microphone array with \SI{4.25}{cm} radius records the audio. Since its publication and use during the 2020 Frederick Jelinek Memorial Summer Workshop, \LibriCSS has become the de facto standard for evaluation and comparison of meeting transcription systems. Those systems typically consist of a separation and enhancement stage, a diarization component and an \gls{ASR} system, where the order of these tasks may vary \cite{Raj2021}.

Since research on meeting scenarios is getting more attention~\cite{yoshioka2019meeting, Horiguchi20, Wang21} and  one data set cannot cater all scenarios, several extensions have been proposed.  As \LibriCSS is merely a test set, the Multi-Speaker Mixture Signal Generator (MMS-MSG) can be used to generate training data for meeting recognition systems: it can be parameterized to take utterances from the \textit{LibriSpeech} corpus to artificially generate meeting data with a desired degree of speech overlap \cite{Landwehr2022}. 
Further, the authors of~\cite{Guan21}  re-recorded \textit{LibriSpeech} data with a microphone array consisting of $40$ synchronized microphones distributed in a room of \SI{110}{m^2} to assess the performance of ad-hoc microphone arrays in a rather large setting. In \cite{Wang21}  spatially distributed asynchronous microphones are used to record a data set named \textit{AdHoc-LibriCSS}. It consists of mini-sessions with either two speakers or five speakers. Each mini-session was recorded by $5$ recording devices, where the placement of the loudspeakers and recording devices were randomly chosen for each session. However, that data set is not publicly available.

We here also focus on an ad-hoc \gls{WASN} scenario. Having in mind a usage scenario where meeting participants use their own smartphones to record the meeting, we record the data with a set of smartphones, whose sampling clocks are not synchronized. The data is further recorded with multi-microphone devices, accounting for a situation, where a multi-channel conference communication system is additionally available for signal capture.

\sloppy
If multiple devices are employed, the sampling rates of each device will slightly differ from the target sampling frequency. This effect, named \gls{SRO},  depends on environmental influences \cite{Walls92}, e.g., voltage supply or temperature, and is time-varying \cite{Gburrek22}. As reported in ~\cite{guggenberger15} the \glspl{SRO} of smartphones and audio devices can be in the range between \SI{-40}{\gls{ppm}} and \SI{416}{\gls{ppm}}, whereby devices from the same vendors have less variation.

Our data set, called \NTLibri, was recorded in two acoustically different rooms, one with a reverberation time of about 200 ms and the other with a reverberation time of about 800 ms. It offers \SI{20}{h} of transcribed audio from $29$ microphones from $9$ different devices. This includes five smartphones, one soundcard and three self-developed smart devices based on Raspberry Pis. It has the same directory structure as \LibriCSS to ease its adoption  for those already working with \LibriCSS. Since playing back the \textit{LibriSpeech} utterances which were also used to create the $60$ \LibriCSS subsets the same \gls{ASR} tools and models can be used to evaluate the data. We complement the data set with ground-truth information about speaker activity for performance analysis purposes.

The data set is intended to be used to conduct research on multi-channel source separation and meeting transcription tasks using asynchronous audio streams. Furthermore, the combination of multi-channel synchronous smart devices with  additional asynchronous single-channel audio streams from smartphones can be studied. Here, approaches which either first estimate and compensate for the \glspl{SRO} \cite{Markovich-Golan2012, miyabe13, Bahari2017} and then separate  the speaker signals as well as approaches which handle the task in an integrated fashion~\cite{Wang21} are conceivable.

The paper is structured as follows: In Sec.~\ref{Sec:Setup} the recording setup and details on the used hardware devices are presented. Then information about the recorded signals and preprocessing steps are summarized in Sec.~\ref{Sec:Signals}. After explaining our reference system in Sec.~\ref{Sec:Baseline} and showing results on the data set in Sec.~\ref{Sec:Experiments} we conclude in Sec.~\ref{Sec:Summary} with a short summary.

\section{Recording Setups}\label{Sec:Setup}

\begin{figure}[htb]
	\centering
	\includegraphics[width=.7\columnwidth]{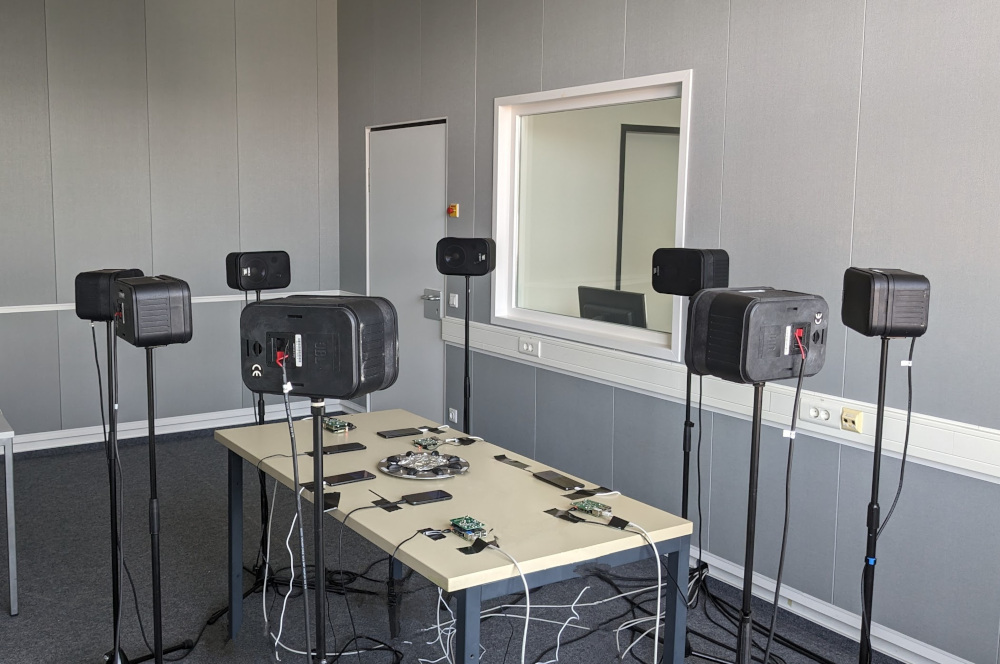}
	\caption{Recording setup of \NTLibriTwo: Eight loudspeakers surrounding a table with multiple recording devices.}
	\label{fig:lab}
\end{figure}

Fig.~\ref{fig:lab} displays the acoustic laboratory room with the recording setup for \NTLibriTwo, i.e., the \NTLibri subset for a room reverberation time of $T_{60} \approx \SI{200}{ms}$. This room has ceiling-high windows on one side of the room with radiators in front of the windows. The ceiling is suspended with mineral fiber boards and the floor is covered with low-pile carpet. A window to the neighboring room, a door and two tables are sound-reflecting elements. The remaining walls are partially covered with a  sound-absorbing surface.

The second room, used for recording \NTLibriEight, is a laboratory room with lightweight walls, a linoleum floor, many window panes and a furnishing consisting of a glass cabinet and tables. This all together leads to an increased reverberation time of $T_{60} \approx \SI{800}{ms}$. Furthermore, the room contained several computers as noise sources.

\begin{figure}[b]
	\centering
	\includegraphics[width=0.8\columnwidth]{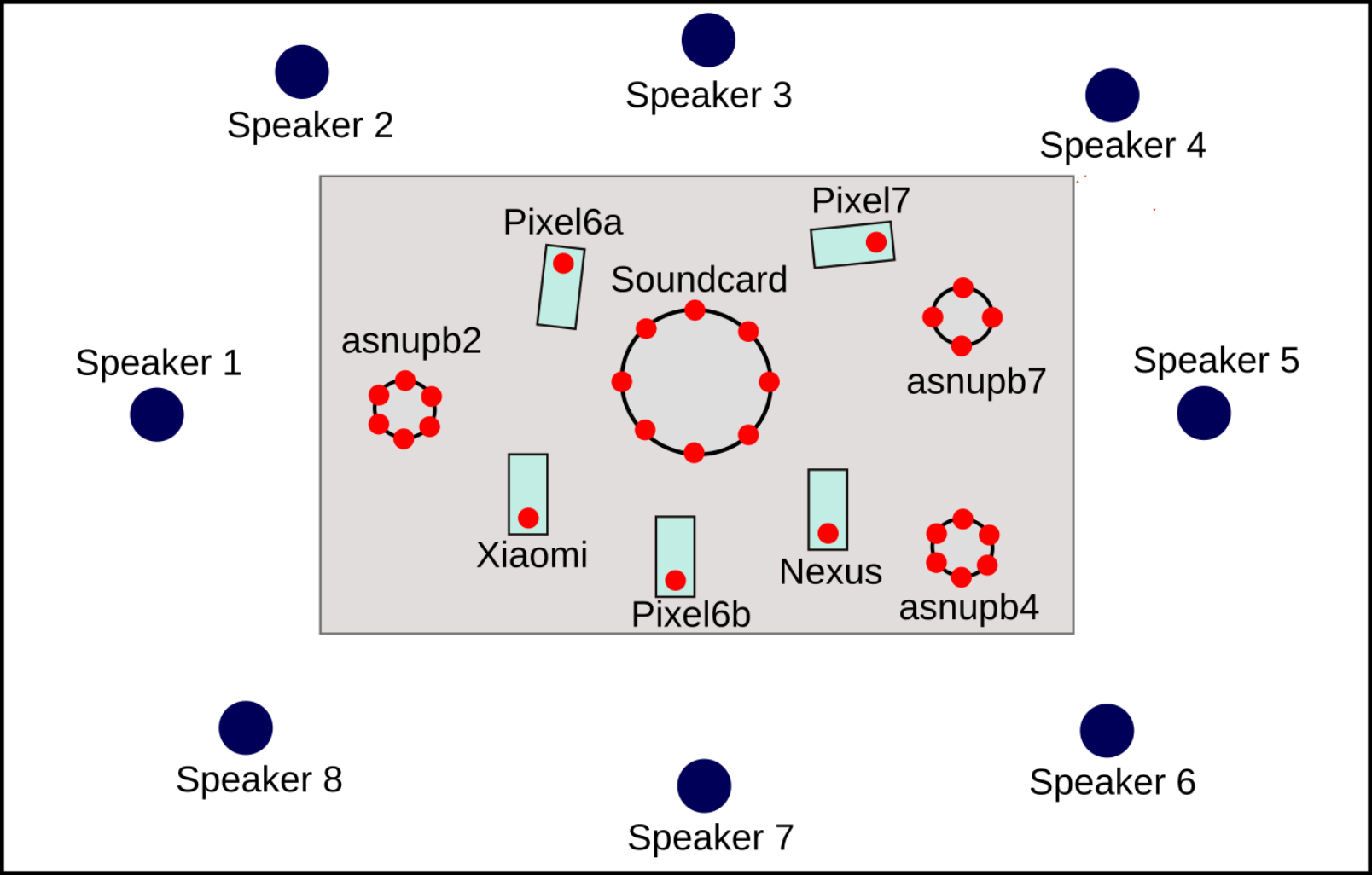}
	\caption{\NTLibriTwo recording and playback devices: Eight loudspeakers surrounding a table with $5$ smartphones, one $8$-channel microphone array (soundcard) and three Raspberry Pi devices. Red dots indicate microphones on devices.}
	\label{fig:scenario}
\end{figure}

A sketch of the device placement for \NTLibriTwo is shown in Fig.~\ref{fig:scenario}. In the center of the table a circular microphone array consisting of eight \textit{AKG C 400 bl} microphones with hypercardioid characteristic is placed. It is connected to the soundcard that plays the  signals on the indicated loudspeaker positions. We used \textit{JBL Control One} loudspeakers that were facing the table. Further, the smartphones and Raspberry Pis are distributed on the table as shown. 

For \NTLibriEight the soundcard array remained in the center of the table, but the positions of the other devices on the table were changed. Position information for all devices and loudspeakers as well as pictures of the setups and rooms can be found in \cite{ZenodoLibriWASN}.

\subsection{Hardware Devices}
The following devices are used to record the data:
\begin{itemize}
	\item Soundcard: Focusrite Scarlett 18i20 (3rd Gen), $8$ channels,  circular microphone array (diameter: \SI{20}{\centi \metre})
	\item Raspberry Pi 4 Model B
	\begin{itemize}
		\item \asnupbTwo \& \asnupbFour: AudioInjector Octo, 6 channel analog frontend, circular microphone array (diameter: \SI{5}{\centi \metre})
		\item \asnupbSeven: Soundcard with 4 channels, adjustable sampling rate at  \gls{ppb} precision, quadratic microphone array (edge length: \SI{5}{\centi \metre})
	\end{itemize}
	\item Android Smartphones
	\begin{itemize}
		\item Xiaomi MI A2 \hspace{1.66cm} {\bf --} Google Pixel 6a ($2 \times$)
		\item LG Group Nexus 4 \hspace{1.1cm} {\bf --} Google Pixel 7 
	\end{itemize}
\end{itemize}
A total of $29$ microphone channels are recorded in a parallel process. The audio playback is handled by the soundcard. See \cite{asnhardware} and \cite{Afifi18} for more details on the 6 channel microphone analog frontend. To distinguish between the two \textit{Pixel 6a} type smartphones we call one ``Pixel 6a'' and the other ``Pixel 6b''.

\subsection{Adjustable Sampling Rate Offset}
\begin{figure}[tb]
	\centering
	\includegraphics[width=0.4\columnwidth]{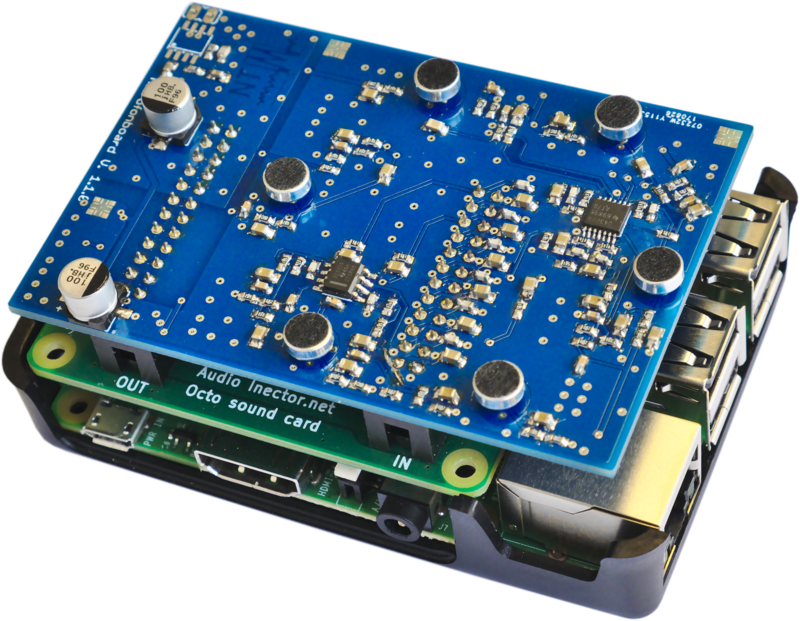}
	\includegraphics[width=0.45\columnwidth]{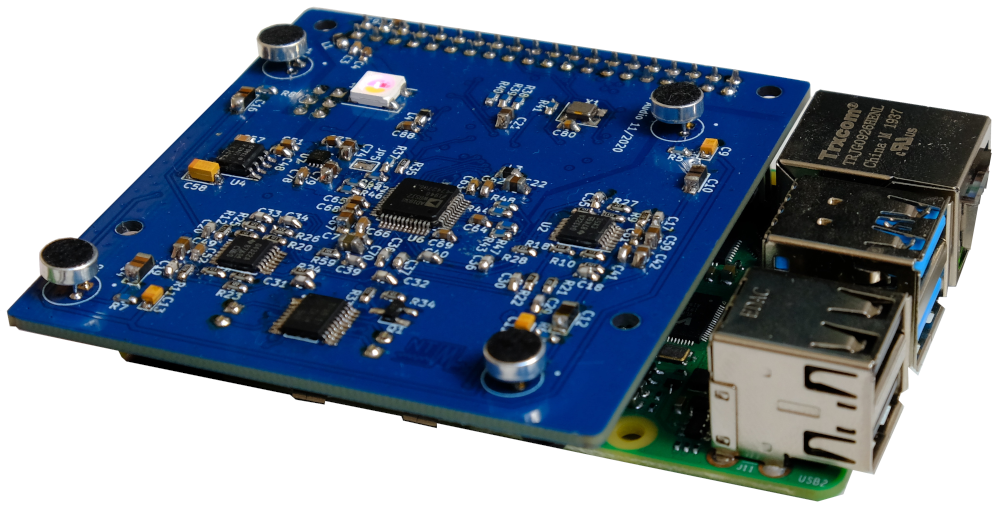}
	\caption{Raspberry Pis with mounted soundcards. Left: AudioInjector Octo with 6 channel microphone analog frontend (\asnupbTwo \& \asnupbFour). Right: \gls{FPGA}-based 4 channel soundcard (\asnupbSeven).}
	\label{fig:rpi}
\end{figure}

\gls{SRO} manipulations in software have high computational costs~\cite{Schmalenstroeer2018} and, furthermore, may attenuate the upper frequencies bands due to the required low-pass filter \cite{Chinaev2018a}. Hence, we developed an audio front-end for Raspberry Pis whose sampling frequency can be adjusted  in hardware at \gls{ppb} granularity (see Fig.~\ref{fig:rpi} right). It uses a \gls{FPGA} for handling the interface communication between hardware and kernel space drivers. Furthermore, it has the ability to compensate for \glspl{SRO} without additional computational requirements with smooth frequency changes in the range of $\pm \SI{1000}{ppm}$ during recording and a lossless recording of all upper frequencies w.r.t. the Nyquist theorem. To this end, on the frontend an Si514 chip (any-frequency \gls{I2C} programmable \gls{xo}) is deployed to generate the sampling signal clock of the \gls{ADC}. 

While the hardware is originally intended to compensate for \glspl{SRO}, we here misuse it to artificially increase the \glspl{SRO}. Hardware bought from the market comes with a random \gls{SRO} and, depending on the individual devices, a certain \gls{SRO} spread can be observed. As reported in~\cite{guggenberger15} typical \gls{SRO} values range between \SI{-40}{\gls{ppm}} and \SI{416}{\gls{ppm}} whereby values exceeding $\pm \SI{100}{ppm}$ are rarely observed. Hence, we set the sampling frequency of \asnupbSeven to \SI{16}{kHz} with an \gls{SRO} of  $-\SI{100}{ppm}$.

Researchers working on the data set can select a subset of devices to meet their requirements w.r.t.\ the \gls{SRO}-range. In Sec.~\ref{Sec:Experiments} some \gls{SRO} measurements are presented.

\subsection{Smartphones \& Streaming}
The Android operating system offers multiple different audio sources for recording. Some of them preprocess the microphone signals, e.g., type \textit{voice recognition}  suppresses environmental noise to a certain extent. 
To prevent signal loss or artifacts on all smartphones the least processed type was selected.

The recorded audio data was streamed wirelessly from each smartphone to a central data acquisition server. Similarly, the Raspberry Pis send their data via a wired connection. All connections are \gls{TCP}-based and have been continuously supervised to work losslessly.


\section{Signals \& Preprocessing} \label{Sec:Signals}
The recordings were performed during the regular operating of our laboratory at the university. As a result they may contain minimal interference from road traffic or impulsive interference from doors.

\subsection{Data Quantization \& Normalization}
The Android smartphones sample the data at \SI{16}{kHz} with a quantization of signed \SI{16}{Bit}. The Raspberry Pis share the same sampling rate with a quantization of signed \SI{32}{Bit} and the soundcard has a sampling rate of \SI{48}{kHz} with \SI{32}{Bit} resolution. The gain of the analog circuits was adjusted so that no clipping occurred during the recordings and all audio stream samples are converted to \SI{16}{Bit} little endian values.

\subsection{Downsampling}
High-quality multi-channel soundcards with native support of a sampling rate of $\SI{16}{kHz}$ are currently not available on the market. Hence, we used a \textit{Focusrite Scarlett 18i20 (3rd Gen)} \gls{USB} soundcard to playback and record the \textit{LibriSpeech} data at \SI{48}{kHz}. To this end, we implemented an up/down-sampling method employing a linear phase, optimal equiripple \gls{FIR} filter with a stop band attenuation of $\SI{50}{dB}$ and a filter length of $256$  taps.

\subsection{High-Pass Filtering}
The recordings of \NTLibriTwo showed low-frequency interference from the water pipes of the heating and an air conditioner from an adjacent lab room, which we removed by high-pass filtering (\SI{3}{dB} cutoff frequency at \SI{75}{Hz}) from all recordings.
After removing the low frequency noise the signals have a \gls{SNR} of approximately $\SI{30}{dB}$ depending on their position and the active source. For example, the Raspberry Pi \textit{asnupb4} has a measured  \gls{SNR} of $\SI{15.25}{dB}$ (in set \textit{OV10, session 3, first speaker}) and after high-pass filtering an \gls{SNR} of $\SI{33.21}{dB}$. The recordings of \NTLibriEight contain more noise from the computers in the  background and no noise from the air conditioning system (e.g., \textit{asnupb4} has an approximate \gls{SNR} of \SI{20}{dB}, a high-pass filter increases it by $\SI{3}{dB}$). Thus, we have taken them unprocessed and leave any noise suppression to future users.

\subsection{Sampling Time Offset Reduction}
The asynchronous devices in the network start recording at unknown points in time. The resulting initial time offset, called \gls{STO}, is not only influenced by quantities that can be controlled, e.g., the selected packet size when interacting with the soundcard and the size of the soundcard data buffer, but also by random quantities that cannot be controlled, e.g., the latency in the network, the packet size of the network and the scheduling of the kernel.

We reduce the \gls{STO} to a technically possible minimum by first starting the recording on all devices and merging the data streams at a central node. After all devices have delivered data, we discard already received data in the queue and start signal recording. This reduces the STO to a size of a few packets. The remaining offset is determined by correlating the signals and then reduced to an approximate range of $\pm \SI{40}{samples}$ during the first \SI{10}{s} of each recording. Note that the described \gls{STO} minimization is done per device so that the intra-device \glspl{TDOA} are maintained.

The  inter-device \glspl{TDOA} correspond to a superposition of the \gls{STO} and the \gls{TDOF}~\cite{Gburrek22}, which is caused by the different distances of the sources to the microphones. 
Thus, the above time offset removal, which forces these \glspl{TDOA} to be close to zero at the beginning of the recordings, does not only remove the \gls{STO} but also manipulates the \gls{TDOF} information.
Since the latter is carrying the source position information source localization based on inter-device \glspl{TDOA}, estimated from the processed signals, cannot be performed.

\section{Reference System} \label{Sec:Baseline}
In addition to the data set we also provide a meeting transcription pipeline as reference/baseline system for future works.
In the experimental section we provide results for this pipeline using single device and multi device setups. 
To enable a coherent processing of signals gathered by different devices, we firstly compensate for \glspl{SRO}.
For this we use the \gls{DWACD} method~\cite{Gburrek22} to estimate the \glspl{SRO} and afterwards compensate for them using the \gls{STFT}-resampling from~\cite{Schmalenstroeer2018}.

In order to extract the single speakers' signals, mask-based beamforming is utilized.
A \gls{cACGMM}~\cite{Ito2016cACGMM} with time-dependent instead of frequency-dependent mixture weights~\cite{Ito2013permutation} is used to estimate a mask for each of the speakers and an additional mask for noise. 
The initialization of the \gls{cACGMM} is based on the idea to divide the meeting into segments consisting of multiple frames, which are clustered afterwards \cite{Boeddeker22}.
For each segment a \gls{SCM} is estimated.
To avoid ambiguities due to speech pauses or overlapping speech, a \mbox{rank-1} approximation of the \glspl{SCM} is conducted. 
In addition to that, the ratio of the largest and the second largest eigenvalue of the  \glspl{SCM} is used as indicator if one speaker is dominant within a segment. 
If this ratio is below a certain threshold, indicating either a speech pause or overlapping speech, the segment is assigned to the noise class.
The segments are clustered based on the similarity of their \glspl{SCM}, which is measured by the the correlation matrix distance from~\cite{Herdin05}. 
By quantizing the similarity measure to zero (not the same speaker) or one (the same speaker) for each segment based on a certain threshold, results in an activity pattern that indicates in which segments the same speaker is active. 
Subsequently, a leader-follower clustering is used to group the segments whose activity patterns intersect most. 
To avoid that the \gls{cACGMM} diverges too much from the initialization, the latter is used as guide for the first \gls{EM} iterations.

\sloppy
The speakers' signals are extracted using a \gls{MVDR} beamformer in the formulation of~\cite{Souden2010MVDR}.
Therefore, the meeting is segmented using the priors of the \gls{cACGMM} as described in~\cite{Boeddeker22}. 
In each segment the \gls{SCM} of the target speaker and the \gls{SCM} of the interference are estimated using the masks obtained from the \gls{cACGMM}. 
The mask used to estimate the interference \gls{SCM} is obtained as sum of all masks except the one of the target speaker.
Finally, a pretrained \gls{ASR} system~\cite{shinji_watanabe_2020_3966501} for \textit{LibriSpeech} from the ESPnet framework~ \cite{watanabe2018espnet} is used to transcribe the separated signals.
This \gls{ASR} system has a transformer architecture and is trained on \textit{LibriSpeech}. On the clean test set of \textit{LibriSpeech}  it achieves a \gls{WER} of \SI{2.7}{\%}.

\section{Experiments}\label{Sec:Experiments}
In the following two aspects of the data set are investigated.
First, the \glspl{SRO} between the different recording devices are analyzed.
Afterwards, the proposed reference system is investigated for varying recording setups.

\subsection{Sampling Rate Offsets}
\begin{table}[t]
	\centering \renewcommand{\arraystretch}{1.3}
	\begin{tabular}{ c c c c c}
		\cline{1-2} \cline{4-5}
		Smartphone & \gls{SRO} && Device & \gls{SRO} \\
		\cline{1-2} \cline{4-5}
		Pixel6a & \SI{16.62}{ppm} && Soundcard & \SI{0.0}{ppm} (Ref.) \\
		Pixel6b & \SI{13.58}{ppm}&& asnupb2 & \SI{-10.35}{ppm}\\
		Pixel7 & \SI{13.73}{ppm}&& asnupb4 & \SI{-23.18}{ppm}\\
		Nexus &  \SI{-0.11}{ppm}&& asnupb7 & \SI{-109.03}{ppm}\\
		Xiaomi &  \SI{-1.44}{ppm}&& & \\
		\cline{1-2} \cline{4-5}
	\end{tabular}
	\caption{Estimated average \glspl{SRO} of \gls{WASN} devices w.r.t.\ soundcard as reference.} 
	\label{tab:dyn_sro}
\end{table}
We estimated the average \gls{SRO} of the hardware device from the audio recordings by employing the \gls{DWACD} method. As shown in Tab.~\ref{tab:dyn_sro} the hardware dependent \glspl{SRO} of the devices are in the range between $\pm \SI{20}{ppm}$ w.r.t.\ the sampling rate of the soundcard. 

Since \cite{guggenberger15} reported \glspl{SRO} of more than $\SI{100}{ppm}$, we decided to enrich the data set with a device that has an artificially higher \gls{SRO}. To this end, we intentionally set the changeable sampling frequency of \textit{asnupb7} to $\SI{15998.4}{kHz}$ relative to its build-in oscillator corresponding to an additional \gls{SRO} of $\SI{-100}{\gls{ppm}}$.
This results in an average \gls{SRO} of $\SI{-109.03}{ppm}$ relative to the soundcard. Note that during the separate recording sessions the device temperatures changed and, thus, the \glspl{SRO} are slightly time-varying.

\begin{figure}[h]
	\centering
	\footnotesize
	\input{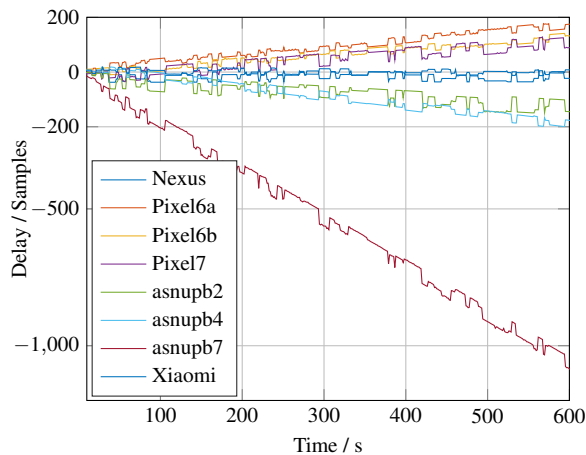}
	\caption{Estimated delay between soundcard and first channel of each device (\NTLibriTwo, OV40, Ses.~9).}
	\label{fig:delay}
\end{figure}
In Fig.~\ref{fig:delay}  the estimated delays between the first channel of the soundcard and the first channel of the other devices are exemplary depicted over time. While the measures described in the last section lead to a delay of nearly zero at the start of the recording, the delays drift over time due to the different \glspl{SRO}. One can also clearly see the abrupt changes in the delay values, that are caused by changing \glspl{TDOF} due to speaker changes.

\begin{table*}[t]
	\begin{adjustbox}{max width=0.99\textwidth}
		\renewcommand{\arraystretch}{1.75}
		\centering
		\begin{tabular}{llcccccccccc} 
			& & & & &\multicolumn{7}{c}{\vspace{-0.6cm} cpWER / \% } \\
			& System information [Device] & \rotatebox{90}{Sync.} & \rotatebox{90}{Devices} & \rotatebox{90}{Channels} & 0L & 0S & OV10 & OV20 & OV30 & OV40 & Avg. \\\hline\hline
			& Clean & - & - & - & 2.90 & 2.58 & 2.58 & 2.43 & 2.51 & 2.31 & 2.52 \\\hline \hline
			\multirow{2}{5ex}{\rotatebox{90}{\footnotesize \LibriCSS}} & \SysA: Segmentation (Oracle act.) & - & 1 & 1 & 4.36 & 4.49 & 10.61 & 18.72 & 27.21 & 35.98 & 18.56 \\\cline{2-12}
			& \SysB: cACGMM \& MVDR  & - & 1 / 1 & 7 / 7 & 3.57 & 3.44 & 3.76 & 4.38 & 5.40 & 5.38 & 4.43 \\\hline \hline
			%
			\multirow{4}{5ex}{\rotatebox{90}{\footnotesize \NTLibriTwo}\: \rotatebox{90}{\footnotesize {$T_{60}\approx \SI{200}{ms}$}}} & \SysA: Segmentation (Oracle act.) [Pixel7] & - & 1 & 1  & 3.69 & 3.47 & 12.21 & 21.84 & 30.73 & 39.95 & 20.59 \\\cline{2-12}
			& \SysB: cACGMM [asnupb4] \& MVDR [asnupb4] & - & 1 / 1 & 6 / 6 & 3.11 & 3.2 & 4.68 & 5.35 & 5.00 & 4.60 & 4.43 \\\cline{2-12}
			& \SysC: cACGMM [asnupb4] \& MVDR [all] & \checkmark & 1 / 9 & 6 / 9 & 3.22 & 3.01 & 4.38 & 5.48 & 3.67 & 4.50 & 4.11 \\\cline{2-12}
			& \SysD:  cACGMM [all] \& MVDR [all] &  \checkmark & 9 / 9 & 9 / 9  & 3.20 & 3.56 & 5.36 & 10.37 & 3.43 & 5.54 & 5.38 \\\hline \hline
			%
			\multirow{4}{5ex}{\rotatebox{90}{\footnotesize \NTLibriEight}\: \rotatebox{90}{\footnotesize {$T_{60}\approx \SI{800}{ms}$}}} & \SysA: Segmentation (Oracle act.) [Pixel7] & - & 1 & 1   & 4.71 & 4.70 & 13.40 & 22.97 & 32.03 & 41.71 & 21.89\\\cline{2-12}
			& \SysB: cACGMM [asnupb4] \& MVDR [asnupb4] & - & 1 / 1 & 6 / 6  & 3.86 & 3.92 & 5.26 & 7.09 & 7.15 & 6.91 & 5.90 \\\cline{2-12}
			& \SysC: cACGMM [asnupb4] \& MVDR [all] &  \checkmark & 1 / 9 & 6 / 9  & 3.87 & 3.55 & 4.14 & 5.33 & 4.94 & 4.59 & 4.47\\\cline{2-12}
			& \SysD:  cACGMM [all] \& MVDR [all] &  \checkmark & 9 / 9 & 9 / 9  & 3.93 & 3.62 & 3.90 & 3.99 & 4.46 & 10.01 & 5.20\\\hline \hline
		\end{tabular}
	\end{adjustbox}
	\caption{Comparison of cpWERs on \LibriCSS and \NTLibri. The cpWER is calculated for the different overlap ratios / silence conditions defined by the \textit{LibriCSS} data set:  no overlap with long silence (0L),  no overlap with short silence (0S) and between \SI{10}{\%} (OV10) and \SI{40}{\%} (OV40) overlap. \textit{Clean} denotes transcribing the original  LibriSpeech utterances used to record \LibriCSS and \NTLibri. \textit{Segmentation} means that oracle activity information is used to cut the unprocessed signal of one channel into blocks, which correspond to single utterances. The first number in the \textit{Devices}/\textit{Channels}-column defines the number of devices/channels used for mask estimation and the second number the number of devices/channels used for beamforming. If signals of multiple devices are jointly aggregated only the first channel of each device is utilized. \vspace{-5mm}}
	\label{tab:wer}
\end{table*}
\subsection{Source Separation}
\Cref{tab:wer} shows a comparison of the \textit{LibriCSS} data set and the two subsets of the \textit{LibriWASN} data set w.r.t.~the performance of the proposed reference meeting transcription system. 
As measure for the meeting transcription performance the \gls{cpWER}~\cite{watanabe20b_chime} is utilized.

In order to get an impression of the acoustic conditions of \textit{LibriWASN} w.r.t. the \gls{ASR} performance, we segmented the unprocessed signal of one channel using oracle activity information for each speaker~(Sys-1).
It becomes obvious that the acoustic conditions (see sessions without overlap: 0L and 0S) of \textit{LibriWASN}$^{200}$ seem to be less challenging w.r.t.~\gls{ASR} than the acoustic conditions of \textit{LibriCSS}.
In contrast, the acoustic conditions of \textit{LibriWASN}$^{800}$ seem to be a little bit more challenging than the acoustic conditions of \textit{LibriCSS} due to greater reverberation and more noise.

Using a single microphone array for mask estimation and beamforming~(Sys-2), it can be seen that similar results can be achieved for \textit{LibriCSS} and \textit{LibriWASN}$^{200}$. 
Again, the performance for \textit{LibriWASN}$^{800}$ is slightly worse due to the more challenging acoustic conditions.
If the masks are still estimated using a single array but all devices are used for beamforming~(Sys-3), the results for both subsets of \textit{LibriWASN} can be improved. 
This especially holds for \textit{LibriWASN}$^{800}$.
Thus, the achieved synchronization based on the \gls{DWACD} method seems to be good enough to enable a coherent signal processing and, therefore, to benefit from the spatial diversity of the distributed recording setup. 

Utilizing all devices for mask estimation and beamforming~(Sys-4), the overall performance of the reference system degrades slightly. 
This is caused by a few sessions with very large \glspl{cpWER}. 
We hypothesize that these outlier results are caused by errors made during the initialization of the \gls{cACGMM}, e.g., by mixing the activity of two speakers. 
However, the good results for some overlap ratios indicate that mask estimation using spatial information can also benefit from spatial diversity of the recording setup.  

Although the results achieved on the \textit{LibriWASN} data set are already decent there is still room for further improvement when comparing the results to the \gls{cpWER} which can be achieved by transcribing the clean utterances from \textit{LibriSpeech} (first result line of \Cref{tab:wer}).
Further, it is to be mentioned that the proposed reference system is based on spatial information, which is a very powerful source of information in a static setup with fixed source positions. 
How to achieve comparable performance with a single-channel system or with moving speakers, remains an open problem.

\section{Summary} \label{Sec:Summary}
We have presented a data set which consists of  re-recordings of the \LibriCSS data set in two different acoustic environments with 9 devices and a total of 29 channels. 
Its intended use is  for diarization and meeting transcription research in ad-hoc wireless acoustic sensor networks, with a focus on synchronization and multi-channel signal processing.

\section*{Download \& Licence}
The data set is available under Creative Commons Attribution 4.0 International License (CC BY 4.0) from Zenodo.
\begin{itemize}
    \item Zenodo data set link: \\ \url{https://zenodo.org/record/7960972}
    \item Scripts for data set handling and code of the reference system: \\ \url{https://github.com/fgnt/libriwasn}   
\end{itemize}

\section*{Acknowledgment}
Funded by the Deutsche Forschungsgemeinschaft (DFG, German Research Foundation) - Project 282835863.

\small
\bibliographystyle{IEEEbib}
\bibliography{refs}


\end{document}

%% file: glossaries.tex
\newacronym{DNN}{DNN}{deep neural network}
\newacronym{CDR}{CDR}{coherent-to-diffuse power ratio}
\newacronym{DoA}{DoA}{direction-of-arrival}
\newacronym{MSE}{MSE}{mean-squared error}
\newacronym{MLP}{MLP}{multilayer perceptron}
\newacronym{CRNN}{CRNN}{convolutional recurrent neural network}
\newacronym[plural=GPs,firstplural=Gaussian processes (GPs)]{GP}{GP}{Gaussian process}
\newacronym{GRU}{GRU}{gated recurrent unit}
\newacronym{RIR}{RIR}{room impulse response}
\newacronym{AWGN}{AWGN}{additive white Gaussian noise}
\newacronym{SNR}{SNR}{signal-to-noise ratio}
\newacronym{STFT}{STFT}{short-time Fourier transform}
\newacronym[plural=PSDs,firstplural=power spectral densities~(PSDs)]{PSD}{PSD}{power sprectral density}
\newacronym{AE}{AE}{absolute error}
\newacronym{MAE}{MAE}{mean-absolute error}
\newacronym{PE}{PE}{position error}
\newacronym{MPE}{MPE}{mean position error}
\newacronym{WASN}{WASN}{wireless acoustic sensor network}
\newacronym{OoR}{OoR}{out-of-range}
\newacronym{ASR}{ASR}{automatic speech recognition}
\newacronym{TDNN}{TDNN}{time delay neural network}
\newacronym{CNN}{CNN}{convolutional neural network}
\newacronym{WLS}{WLS}{weighted least squares}
\newacronym{LS}{LS}{least squares}
\newacronym{RANSAC}{RANSAC}{random sample consensus}

\newacronym{GARDE}{GARDE}{\textbf{G}eometry c\textbf{A}libration f\textbf{R}om \textbf{D}istance \textbf{E}stimates}
\newacronym{MDS}{MDS}{Multi Dimensional Scaling}

\newacronym{CWLS}{CWLS}{constrained weighted least squares}
\newacronym{CRLB}{CRLB}{Cramer-Rao lower bound}
\newacronym{RMSE}{RMSE}{root mean square error}

\newacronym{CDF}{CDF}{Cumulative distribution function}
\newacronym{CDF2}{CDF}{Cumulative distribution function}

\newacronym{ABEL}{ABEL}{averaged burst error length}
\newacronym{ACD}{ACD}{average coherence drift}
\newacronym{ADC}{ADC}{analog-digital converter}
\newacronym{APT}{APT}{averaged processing time}
\newacronym{ASN}{ASN}{acoustic sensor network}
\newacronym{ASNs}{ASNs}{acoustic sensor networks}
\newacronym{ASRC}{ASRC}{arbitrary sampling rate conversion}
\newacronym{ATD}{ATD}{time drift}
\newacronym{ATS}{ATS}{accumulating time shift}
\newacronym{BI}{BI}{band-limited interpolation}
\newacronym{BSS}{BSS}{blind source separation}
\newacronym{CCF}{CCF}{cross-correlation function}
\newacronym{CCF-2}{CCF-2}{secondary \gls{CCF}}
\newacronym{CD}{CD}{coherence drift}
\newacronym{CFM}{CFM}{coherence function maximization}
\newacronym{CM}{CM}{correlation maximization}
\newacronym{CSD}{CSD}{cross-spectral density}
\newacronym{CSD-2}{CSD-2}{secondary \gls{CSD}}
\newacronym{CTC}{CTC}{continuous-time conversion}
\newacronym{DCML}{DCML}{\gls{ML} method for dynamic conditions}
\newacronym{DB}{DB}{Data base}
\newacronym{DD}{DD}{digital-to-digital}
\newacronym{DDC}{DDC}{digital-to-digital converter}
\newacronym{DSP}{DSP}{digital signal processing}
\newacronym{DTC}{DTC}{discrete-time conversion}
\newacronym{DXCP}{DXCP}{double-cross-correlation processor}
\newacronym{DXCPPhaT}{DXCP-PhaT}{\gls{DXCP} with phase transform}
\newacronym{ECS}{ECS}{energy correlation score}
\newacronym{FCF}{FCF}{filter correlation function}
\newacronym{FF}{FF}{free-field}
\newacronym{FFT}{FFT}{fast Fourier transform}
\newacronym{FFT-DXCP}{FFT-DXCP}{\gls{FFT} domain \gls{DXCP}}
\newacronym{FO}{FO}{frame-oriented}
\newacronym{GCC}{GCC}{generalized cross-correlation}
\newacronym{GCPSD}{GCPSD}{generalized cross power sprectral density}
\newacronym{GCCF}{GCCF}{generalized \gls{CCF}}
\newacronym{GCCPhaT}{GCC-PhaT}{generalized cross-correlation with phase transform}
\newacronym{GCSD}{GCSD}{generalized cross-spectral density}
\newacronym{GE}{GE}{Gilbert-Elliott}
\newacronym{IBI}{IBI}{iterative \gls{BI}}
\newacronym{ICA}{ICA}{independent component analysis}
\newacronym{IML}{IML}{iterative \gls{ML}}
\newacronym{IR}{IR}{impulse response}
\newacronym{IIR}{IIR}{infinite impulse response}
\newacronym{IFFT}{IFFT}{inverse \gls{FFT}}
\newacronym{ISTFT}{ISTFT}{inverse \gls{STFT}}
\newacronym{LCD}{LCD}{least-squares coherence drift}
\newacronym{LPD}{LPD}{linear-phase drift}
\newacronym{LTI}{LTI}{linear time-invariant}
\newacronym{LTV}{LTV}{linear time-variant}
\newacronym{LTVIR}{LTV-IR}{linear time-variant impulse response}
\newacronym{ML}{ML}{maximum likelihood}
\newacronym{MS}{MS}{multi-stage}
\newacronym{MSC}{MSC}{magnitude-squared coherence}
\newacronym{OML}{OML}{optimized \gls{ML}}
\newacronym{OR}{OR}{outlier removal}
\newacronym{PHAT}{PHAT}{phase transform}
\newacronym{PL}{PL}{packet loss}
\newacronym{PolyFar}{PolyFar}{polyphase-Farrow}
\newacronym{ppb}{ppb}{part per billion}
\newacronym{ppm}{ppm}{parts \ per \ million}

\newacronym{RBI}{RBI}{recursive band-limited interpolation}
\newacronym{RF}{RF}{realtime factor}
\newacronym{RTP}{RTP}{Real-time Transport Protocol}
\newacronym{RTCP}{RTCP}{Real-time Transport Control Protocol}

\newacronym{SCOT}{SCOT}{smoothed coherence transform}
\newacronym{SINR}{SINR}{signal-to-interpolation-noise ratio}
\newacronym{SO}{SO}{sample-oriented}
\newacronym{SPIB}{SPIB}{signal processing information base}
\newacronym{SPL}{SPL}{signal packet loss}
\newacronym{SRC}{SRC}{sampling rate conversion}
\newacronym{SRO}{SRO}{sampling rate offset}
\newacronym{STO}{STO}{sampling time offset}
\newacronym{TD}{TD}{time domain}
\newacronym{TDDXCP}{TD-DXCP}{time domain \gls{DXCP}}
\newacronym{TDD}{TDD}{time-delay difference}
\newacronym{UDP}{UDP}{User Datagram Protocol}
\newacronym{VAD}{VAD}{activity detection}
\newacronym{SAD}{SAD}{sound activity detection}
\newacronym{WACD}{WACD}{weighted average coherence drift}
\newacronym{DWACD}{DWACD}{dynamic weighted average coherence drift}
\newacronym{WG}{WG}{weighting}
\newacronym{WLCD}{WLCD}{weighted \gls{LCD}}
\newacronym{WASNs}{WASNs}{wireless acoustic sensor networks}
\newacronym{WCP}{WCP}{wideband correlation processor}
\newacronym{XC}{XC}{cross-correlation}
\newacronym{FIR}{FIR}{finite impulse response}
\newacronym{AGC}{AGC}{automatic gain control}

\newacronym{TXCO}{TXCO}{temperature compensated crystal oscillator}

\newacronym{OU}{OU}{Ornstein-Uhlenbeck}
\newacronym{CSS}{CSS}{continuous speech separation}

\newacronym{WER}{WER}{word error rate}
\newacronym{cpWER}{cpWER}{concatenated minimum-permutation word error rate}
\newacronym{cACGMM}{cACGMM}{complex Angular Central Gaussian Mixture Model}

\newacronym[longplural={time differences of arrival}]{TDOA}{TDOA}{time difference of arrival}
\newacronym[longplural={time differences of flight}]{TDOF}{TDOF}{time difference of flight}
\newacronym{GSS}{GSS}{guided source separation}
\newacronym{SCM}{SCM}{spatial covariance matrix}
\newacronym{EM}{EM}{expectation maximization}
\newacronym{MVDR}{MVDR}{minimum variance distortionless response}

\newacronym{FPGA}{FPGA}{field programmable gate array}
\newacronym{I2C}{$I^2C$}{inter-integrated circuit}
\newacronym{xo}{xo}{crystal oscillator}
\newacronym{USB}{USB}{universal serial bus}
\newacronym{TCP}{TCP}{transmission control protocol}

%% file: main.bbl
\begin{thebibliography}{10}

\bibitem{20_Chen_LibriCSS}
Zhuo Chen, Takuya Yoshioka, Liang Lu, Tianyan Zhou, Zhong Meng, Yi~Luo, Jian
  Wu, Xiong Xiao, and Jinyu Li,
\newblock ``Continuous speech separation: Dataset and analysis,''
\newblock in {\em IEEE International Conference on Acoustics, Speech and Signal
  Processing (ICASSP)}, 2020.

\bibitem{Libri15}
Vassil Panayotov, Guoguo Chen, Daniel Povey, and Sanjeev Khudanpur,
\newblock ``Librispeech: An {ASR} corpus based on public domain audio books,''
\newblock in {\em IEEE International Conference on Acoustics, Speech and Signal
  Processing (ICASSP)}, 2015.

\bibitem{Raj2021}
Desh Raj, Pavel Denisov, and Zhuo {Chen et al.},
\newblock ``Integration of speech separation, diarization, and recognition for
  multi-speaker meetings: System description, comparison, and analysis,''
\newblock in {\em 2021 IEEE Spoken Language Technology Workshop (SLT)}, 2021,
  pp. 897--904.

\bibitem{yoshioka2019meeting}
Takuya Yoshioka, Dimitrios Dimitriadis, Andreas Stolcke, William Hinthorn, Zhuo
  Chen, Michael Zeng, and Xuedong Huang,
\newblock ``Meeting transcription using asynchronous distant microphones,''
\newblock in {\em Proc. Interspeech}, September 2019.

\bibitem{Horiguchi20}
Shota Horiguchi, Yusuke Fujita, and Kenji Nagamatsu,
\newblock ``Utterance-wise meeting transcription system using asynchronous
  distributed microphones,''
\newblock in {\em Interspeech}, 2020.

\bibitem{Wang21}
Dongmei Wang, Takuya Yoshioka, Zhuo Chen, Xiaofei Wang, Tianyan Zhou, and Zhong
  Meng,
\newblock ``Continuous speech separation with ad hoc microphone arrays,''
\newblock in {\em European Signal Processing Conference (EUSIPCO)}, 2021.

\bibitem{Landwehr2022}
Tobias Cord-Landwehr, Thilo von Neumann, Christoph Boeddeker, and Reinhold
  Haeb-Umbach,
\newblock ``{MMS-MSG}: A multi-purpose multi-speaker mixture signal
  generator,''
\newblock in {\em International Workshop on Acoustic Signal Enhancement
  (IWAENC)}, 2022.

\bibitem{Guan21}
Shanzheng Guan, Shupei Liu, and Junqi {Chen et al.},
\newblock ``Libri-adhoc40: A dataset collected from synchronized ad-hoc
  microphone arrays,''
\newblock in {\em 2021 Asia-Pacific Signal and Information Processing
  Association Annual Summit and Conference (APSIPA ASC)}, 2021, pp. 1116--1120.

\bibitem{Walls92}
Fred~L. Walls and Jean-Jacques Gagnepain,
\newblock ``Environmental sensitivities of quartz oscillators,''
\newblock {\em IEEE transactions on ultrasonics, ferroelectrics, and frequency
  control}, vol. 39, pp. 241--9, 02 1992.

\bibitem{Gburrek22}
Tobias Gburrek, Joerg Schmalenstroeer, and Reinhold Haeb-Umbach,
\newblock ``On synchronization of wireless acoustic sensor networks in the
  presence of time-varying sampling rate offsets and speaker changes,''
\newblock in {\em Proc. International Conference on Acoustics, Speech and
  Signal Processing (ICASSP)}, 2022.

\bibitem{guggenberger15}
Mario Guggenberger, Mathias Lux, and Laszlo B{\"o}sz{\"o}rmenyi,
\newblock ``An analysis of time drift in hand-held recording devices,''
\newblock in {\em MultiMedia Modeling}. 2015, pp. 203--213, Springer
  International Publishing.

\bibitem{Markovich-Golan2012}
Shmulik Markovich-Golan, Sharon Gannot, and Israel Cohen,
\newblock ``Blind sampling rate offset estimation and compensation in wireless
  acoustic sensor networks with application to beamforming,''
\newblock in {\em Proc. International Workshop on Acoustic Echo and Noise
  Control (IWAENC)}, 2012.

\bibitem{miyabe13}
Shigeki Miyabe, Nobutaka Ono, and Shoji Makino,
\newblock ``Blind compensation of inter-channel sampling frequency mismatch
  with maximum likelihood estimation in stft domain,''
\newblock in {\em Proc. IEEE International Conference on Acoustics, Speech and
  Signal Processing (ICASSP)}, 2013, pp. 674--678.

\bibitem{Bahari2017}
Mohamad~Hasan Bahari, Alexander Bertrand, and Marc Moonen,
\newblock ``Blind sampling rate offset estimation for wireless acoustic sensor
  networks through weighted least-squares coherence drift estimation,''
\newblock {\em IEEE/ACM Transactions on Audio, Speech, and Language
  Processing}, vol. 25, no. 3, pp. 674--686, 2017.

\bibitem{ZenodoLibriWASN}
Joerg Schmalenstroeer, Tobias Gburrek, and Reinhold Haeb-Umbach,
\newblock ``{Zenodo: LibriWASN data set (open access)},''
  https://zenodo.org/record/7960972, 2023.

\bibitem{asnhardware}
{DFG FOR 2457},
\newblock ``Homepage acoustic sensor networks project - open hardware,''
  https://upb.de/asn/hardware, 2023.

\bibitem{Afifi18}
Haitam Afifi, Joerg Schmalenstroeer, Joerg Ullmann, Reinhold Haeb-Umbach, and
  Holger Karl,
\newblock ``{MARVELO} - a framework for signal processing in wireless acoustic
  sensor networks,''
\newblock in {\em Speech Communication; 13th ITG-Symposium}, 2018, pp. 1--5.

\bibitem{Schmalenstroeer2018}
Joerg Schmalenstroeer and Reinhold Haeb-Umbach,
\newblock ``Efficient sampling rate offset compensation - an overlap-save based
  approach,''
\newblock in {\em Proc. European Signal Processing Conference (EUSIPCO)}, 2018.

\bibitem{Chinaev2018a}
Aleksej Chinaev, Gerald Enzner, and Joerg Schmalenstroeer,
\newblock ``Fast and accurate audio resampling for acoustic sensor networks by
  polyphase-{F}arrow filters with {FFT} realization,''
\newblock in {\em Proc. of ITG Fachtagung Sprachkommunikation (Speech
  Communications)}, Oct. 2018.

\bibitem{Ito2016cACGMM}
Nobutaka Ito, Shoko Araki, and Tomohiro Nakatani,
\newblock ``Complex angular central gaussian mixture model for directional
  statistics in mask-based microphone array signal processing,''
\newblock in {\em Proc. European Signal Processing Conference (EUSIPCO)}, 2016.

\bibitem{Ito2013permutation}
Nobutaka Ito, Shoko Araki, and Tomohiro Nakatani,
\newblock ``Permutation-free convolutive blind source separation via full-band
  clustering based on frequency-independent source presence priors,''
\newblock in {\em Proc. IEEE International Conference on Acoustics, Speech and
  Signal Processing (ICASSP)}, 2013.

\bibitem{Boeddeker22}
Christoph Boeddeker, Tobias Cord-Landwehr, Thilo von Neumann, and Reinhold
  Haeb-Umbach,
\newblock ``An initialization scheme for meeting separation with spatial
  mixture models,''
\newblock in {\em Proc. INTERSPEECH}, 2022.

\bibitem{Herdin05}
M.~Herdin, N.~Czink, H.~Ozcelik, and E.~Bonek,
\newblock ``Correlation matrix distance, a meaningful measure for evaluation of
  non-stationary {MIMO} channels,''
\newblock in {\em 2005 IEEE 61st Vehicular Technology Conference}, 2005,
  vol.~1, pp. 136--140 Vol. 1.

\bibitem{Souden2010MVDR}
Mehrez Souden, Jacob Benesty, and Sofi{\`e}ne Affes,
\newblock ``On optimal frequency-domain multichannel linear filtering for noise
  reduction,''
\newblock {\em IEEE Transactions on audio, speech, and language processing},
  vol. 18, no. 2, pp. 260--276, 2010.

\bibitem{shinji_watanabe_2020_3966501}
Shinji Watanabe,
\newblock ``{ESPnet2 pretrained automatic speech recognition model,
  https://doi.org/10.5281/zenodo.3966501},'' July 2020.

\bibitem{watanabe2018espnet}
Shinji Watanabe, Takaaki Hori, and Shigeki {Karita et al.},
\newblock ``{ESPnet}: End-to-end speech processing toolkit,''
\newblock in {\em Proc. INTERSPEECH}, 2018.

\bibitem{watanabe20b_chime}
Shinji Watanabe, Michael Mandel, and Jon {Barker et al.},
\newblock ``{CHiME-6 Challenge: Tackling Multispeaker Speech Recognition for
  Unsegmented Recordings},''
\newblock in {\em Proc. 6th International Workshop on Speech Processing in
  Everyday Environments (CHiME 2020)}, 2020, pp. 1--7.

\end{thebibliography}
